\documentclass[aps,final,notitlepage,oneside,nobibnotes,nofootinbib,
superscriptaddress,noshowpacs,centertags]{revtex4-1}
\usepackage{graphicx}
\usepackage{latexsym}
\usepackage{amssymb}
\usepackage{amsmath}
\usepackage{float}
\usepackage{mathtools}
\usepackage{pxfonts}
\usepackage{xcolor}

\begin{document}
\title{Generalized Einstein--Rosen bridge inside black holes}
\author{Vyacheslav I. Dokuchaev}\thanks{e-mail: dokuchaev@inr.ac.ru}
\affiliation{Institute for Nuclear Research of the Russian Academy of Sciences \\
60th October Anniversary Prospect 7a, 117312 Moscow, Russia}
\author{Konstantin E. Prokopev}\thanks{e-mail: k.prokopev@minus.inr.ac.ru}
\affiliation{Institute for Nuclear Research of the Russian Academy of Sciences \\
60th October Anniversary Prospect 7a, 117312 Moscow, Russia}
\begin{abstract}
We generalize the notion of Einstein--Rosen bridge by defining it as a space-like connection  between two universes with regions of asymptotically minkowskian space-time infinities. The corresponding symmetry and asymmetry properties of the generalized Einstein--Rosen bridge are considered at the cases of Reissner--Nordstr\"om and Kerr metrics. We elucidate the  versatility of intriguing  symmetry and asymmetry phenomena outside and inside black holes. For description of the test particle (planet and photon) motion it is used the Kerr-Newman metric of the rotating and electrically charged black hole. In particular, it is demonstrated the symmetry and asymmetry of the one-way Einstein--Rosen bridge inside black hole toward and through the plethora of endless and infinite universes.
\end{abstract}
\maketitle 

\section{Introduction}
\label{intro}

In this paper we generalize the notion of Einstein--Rosen bridge by defining it as a space-ilke connection between two universes with regions of asymptotically minkowskian space-time infinities. The corresponding symmetry and asymmetry properties of the generalized Einstein--Rosen bridge are considered at the cases of Reissner--Nordstr\"om and Kerr metrics. We elucidate the versatility of intriguing symmetry and asymmetry phenomena outside and inside black holes. For description of the test particle (planet and photon) motion it is used the Kerr-Newman metric of the rotating and electrically charged black hole. It is demonstrated the symmetry and asymmetry of the one-way Einstein--Rosen bridge inside black hole toward and through the plethora of endless and infinite universes.

It seems that the original idea of an infinite series of bridges between universes in the Kerr metric belongs to Boyer and Lindquist \cite{BoyerLindquist}. The Reissner-–Nordström and Kerr one-way bridge is discussed in Chapter 6.5 of Carroll’s textbook \cite{Carroll} and also in Chapters 3.5
and 4.4 of Ullmann’s textbook \cite{Ullmann}. The last book also points to the physical obstacles to the existence of such a bridge between universes, which can be associated with various types of instabilities (including quantum ones), which are discussed for example in \cite{Giirsel,SimpsonPenrose} and more modern attempts \cite{DeMott}. However, the problem still remains open. Recently symmetrical geodesic motion, bound and unbound orbits and the possibility of passing through the Reissner-–Nordström and Kerr bridge are also analyzed in \cite{Abramson} and \cite{DysonMeent} respectively.

\section{Kerr-Newman metric for rotating and electrically charged black hole}

The famous Kerr--Newman metric or geometry (see, e.\, g., \cite{Kerr,Carter68,novthorne73,mtw,Chandra}), which is the exact solution of Einstein's equations \cite{Penrose,Bardeen73,Choquet-Bruhat, Wald,hawkingellis,BPT,bch} for a rotating and electrically black hole, is: 
\begin{equation}
	ds^2=-\frac{\Delta}{\Sigma}[dt-a\sin^2\theta d\varphi]^2\!+\!\frac{\sin^2\theta}{\Sigma}[(r^2\!+\!a^2)d\varphi -adt^2]^2\! +\!\frac{\Sigma}{\Delta}dr^2+\!\Sigma d\theta^2\!,
	\label{metric}
\end{equation}
where $(r,\theta,\phi)$ are spherical coordinates and $t$ is the time of static distant observer at the asymptotically radial infinity. In this metric 
\begin{equation}
	\Delta = r^2-2Mr+a^2+q^2, \quad \Sigma= r^2+a^2\cos^2\theta,  	\label{Delta}
\end{equation}
$M$ \ --- \ black hole mass, $q$ \ --- \ black hole electric charge, $a= J/M$ \ --- \ specific black hole angular momentum (spin). The two roots of equation $\Delta=0$ are \ $r_+=1+\sqrt{M^2-a^2-q^2}$ \ --- \ the black hole event horizon and $r_-=1-\sqrt{M^2-a^2-q^2}$ \ --- \ the black hole Cauchy horizon.

For simplification of equation and presentation of Figures we will often use  units $G = 1$, $c = 1$, $M = 1$ and dimensionless radius $r\Rightarrow r/M$ and dimensionless time $t\Rightarrow t/M$.

In the Kerr-Newman metric there the following integrals of motion for test particles \cite{Carter68}: $\mu$ --- particle mass, $E$ --- particle total energy, $L$ --- particle azimuthal angular moment and $Q$ --- the Carter constant, related with the non-equatorial particle motion. The corresponding geodesic equations for test particle motion in the Kerr-Newman metric in the differential form are: 
\begin{eqnarray}
	\Sigma\frac{dr}{d\tau} &= & \sqrt{R}, \label{dr} \\
	\Sigma\frac{d\theta}{d\tau} &= & \sqrt{\Theta}, \label{dtheta} \\
	\Sigma\frac{d\varphi}{d\tau} &= & -a(aE-\frac{L}{\sin^2\theta})+\sin^2\frac{a}{\Delta}, \label{dvarphi} \\
	\Sigma\frac{dt}{d\tau} &= & -a(aE\sin^2-L)+(r^2+a^2)\frac{P}{\Delta}. \label{dt}
\end{eqnarray}
Here 
\begin{equation}
	P=E(r^2+a^2)-a L+
	\epsilon qr, 	\label{Vtheta} 
\end{equation}
$\tau$ --- the proper time of a test massive particle or an affine parameter along the trajectory of a massless particle ($\mu=0$) like photon. Respectively, the effective radial potential $R(r)$ is
\begin{equation}
	R(r) = P^2-\Delta[\mu^2r^2+(L-aE)^2+Q], 	\label{Rr} 
\end{equation}
and the effective polar potential $\Theta(\theta)$ is 
\begin{equation}
	\Theta(\theta) = Q-\cos^2\theta[a^2(\mu^2-E^2)+L^2\sin^{-2}\theta]. 	\label{Vtheta} 
\end{equation}

Trajectories of massive particles ($\mu\neq0$) depend on three  parameters: $\gamma=E/\mu$, $\lambda=L/\mu$ and $Q/\mu^2$. Meantime, trajectories of massless particles like photons (null geodesics) depend on two parameters: $\lambda$ and $Q$.

The nontrivial specific feature of the rotating Kerr black hole ($a\neq0$) is the existence of so-called {\it ergosphere} \cite{mtw,Chandra,Wald,BPT,bch} with the outer boundary 
\begin{equation}\label{ES}
	r_{\rm ES}(\theta) =1+\sqrt{1-q^2-a^2\cos^2\theta}.
\end{equation}
Inside the ergosphere any test object is dragged into insuperable rotation around black hole with infinite azimuthal winding by approaching the black hole event horizon. Note that the winding effect was discussed
also in \cite{DysonMeent,Kling}

In the following Sections we will describe the symmetry and asymmetry of test object motion in the gravitational field of the Kerr-Newman black hole. We use equations of motion in the Kerr--Newman metric (\ref{dr})--(\ref{dt}) in our analytic and numerical calculations of test particle geodesic trajectories \cite{Babichev13,Dokuch14, FizLab, DokEr15,doknaz17, doknaz18b, doknaz19, doknazsm19,doknaz19b, dokuch19,doknaz18c}.

\section{One-way Einstein--Rosen bridge inside black  hole}

\begin{figure}[H]
	\centering
	\includegraphics[angle=0,width=0.5\textwidth]{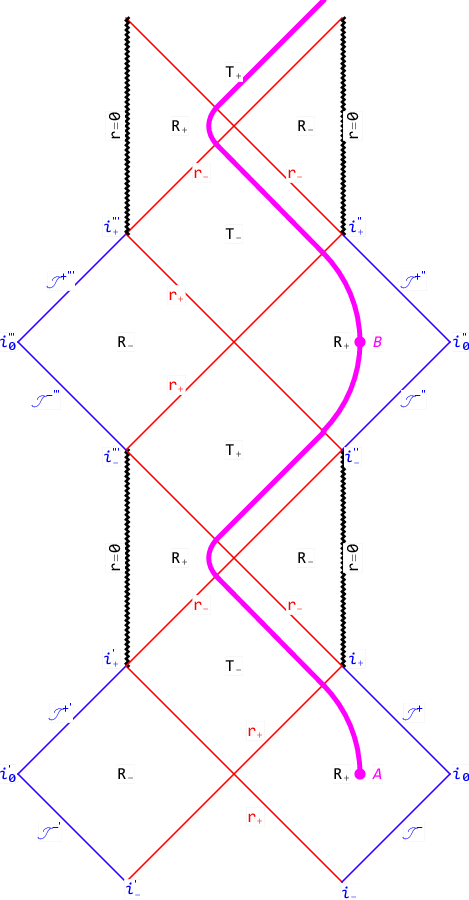} 
	\caption{Carter--Penrose diagram for the spherically symmetric Reissner--Nordstr\"om black hole with electric charge $q=0.99$. The spaceship starts from the point  ${\color{magenta}A}$ at $R_+$-region toward its multi-planetary future inside the black hole. Astronauts are planning to use the Einstein--Rosen bridge {\color{magenta}magenta} curve) and intersect both the black hole event horizon $r_+$ and Cauchy horizon $r_-$ at finite their proper time. After appearing near the black hole singularity at $r=0$, the spaceship uses its powerful engines to change the direction of motion and escape the tidal destruction at small radii. In result, the voyage is happily finishing at point $B$ (may be at the Earth-like planet) in another infinite universe. The symmetry is in possibility to repeat the complete route of this voyage staring from the point  ${\color{magenta}B}$ but only in the forward direction in time toward another multi-planetary future. It is impossible to return the native Earth due to impossibility of any motion beyond the light cone. This is the motion {\it asymmetry} on one-way Einstein--Rosen bridge inside black hole.} 
	\label{fig1}
\end{figure}

\begin{figure}
	\centering
	\includegraphics[angle=0,width=0.75\textwidth]{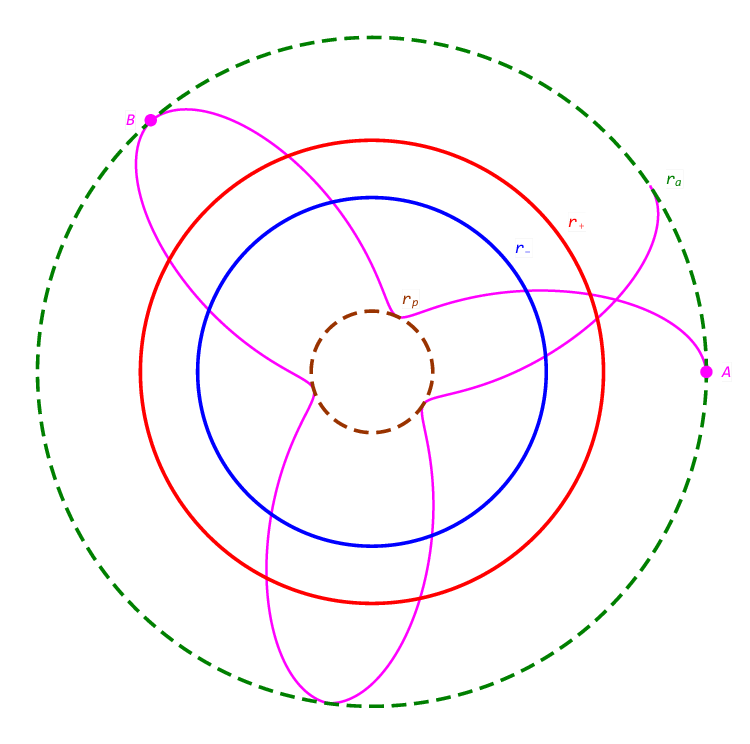} 
	\caption{$2D$ presentation of the voyage through the Reissner--Nordstr\"om black hole interiors by using the Einstein--Rosen bridge. This picture is geometrically absolutely {\it symmetric} or, in other words, it is nicely {\it symmetric}. At the same time, this picture is misleading and physically controversial: Indeed. the voyage is starting at apogee $r_a$ from the position at point ${\color{magenta}A}$, then reach the perigee $r_p$ and return the  apogee $r_a$ at the point $B$ for a finite proper time, demonstrating the absolute geometric {\it symmetry}. Meanwhile, there is a crucial hitch: this apogee $r_a$ at the point ${\color{magenta}B}$ is not in the native universe, but in the other quite distant universe, as it is clearly viewed at the Carter--Penrose diagram at Fig.~\ref{fig1}. The apogee $r_a$ and perigee $r_p$ radii are shown by dashed circles. Respectively, the event radii of event horizon $r_+$ and Cauchy horizon $r_-$ are shown by solid circles. The {\color{magenta}magenta} curve here and at the Fig.~\ref{fig4} is numerically calculated geodesic trajectory with using equations of motion (\ref{dr})--(\ref{dt}) for massive test particles ($\mu\neq0$).}
	\label{fig2}
\end{figure}

\begin{figure}
	\centering
	\includegraphics[angle=0,width=0.85\textwidth]{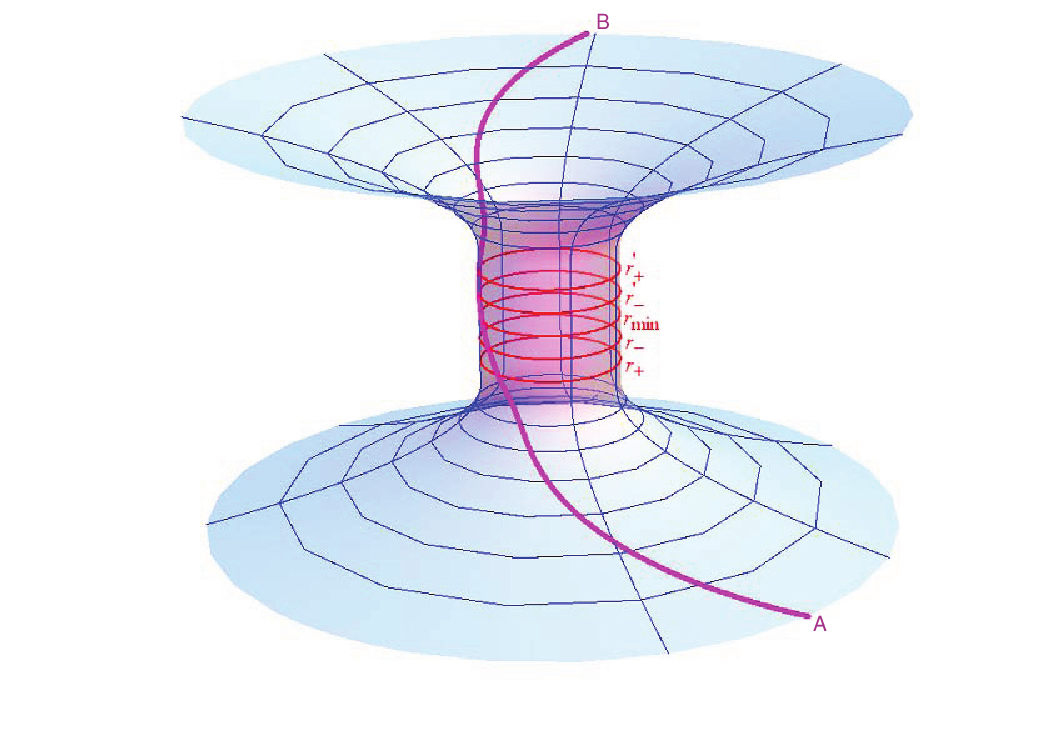} 
	\caption{Embedding diagram for the voyage through black hole interiors by using the Einstein--Rosen bridge. This bridge connects two asymptotically flat universes like wormhole tunnel, but with the only one-way motion from the initial point ${\color{magenta}A}$ to the final point ${\color{magenta}B}$. The geometrical {\it symmetry} of this embedding diagram is deceptive. In fact, this embedding diagram demonstrate the {\it asymmetric} space-time origin of the one-way Einstein--Rosen bridge (remember about loss-cone).}
	\label{fig3}
\end{figure}

\begin{figure}
	\centering
	\includegraphics[angle=0,width=0.65\textwidth]{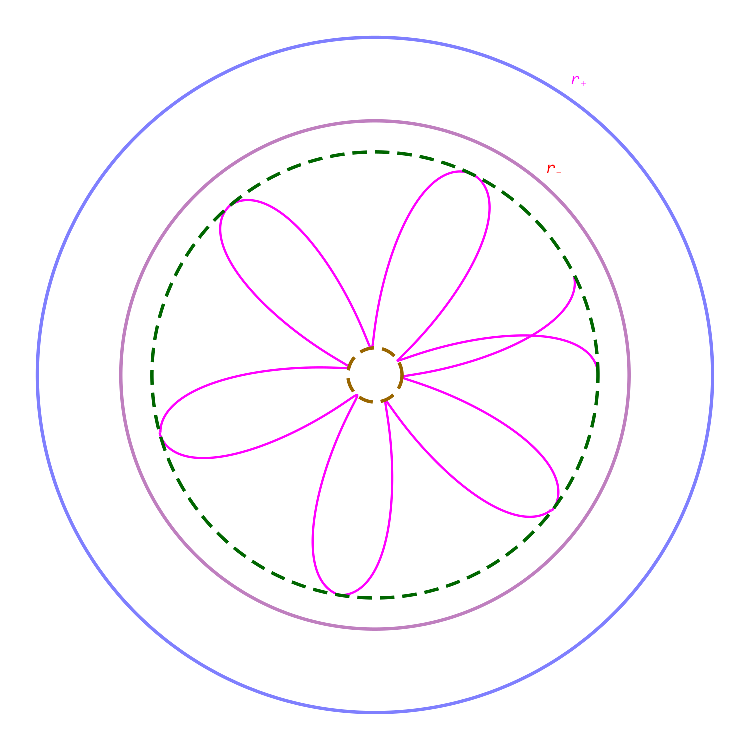} 
	\caption{Both the geometrical and physical completely {\it symmetric} picture of the periodic orbital motion of the test planet or spaceship around the central singularity of the Reissner--Nordstr\"om black hole inside the Cauchy horizon $r_-$. The {\it asymmetric} Reissner--Nordstr\"om bridge is only needed for penetration into this very exotic region at $0<r<r_-$, where exist the nearly stable periodic orbits for test particles, which are very similar to the periodic orbits outside the black hole event horizon $r_+$. The apogee $r_a$ and perigee $r_p$ radii are shown by dashed circles. Respectively, the event radii of event horizon $r_+$ and Cauchy horizon $r_-$ are shown by solid circles.}
	\label{fig4}
\end{figure}

\begin{figure}
	\centering
	\includegraphics[angle=0,width=0.57\textwidth]{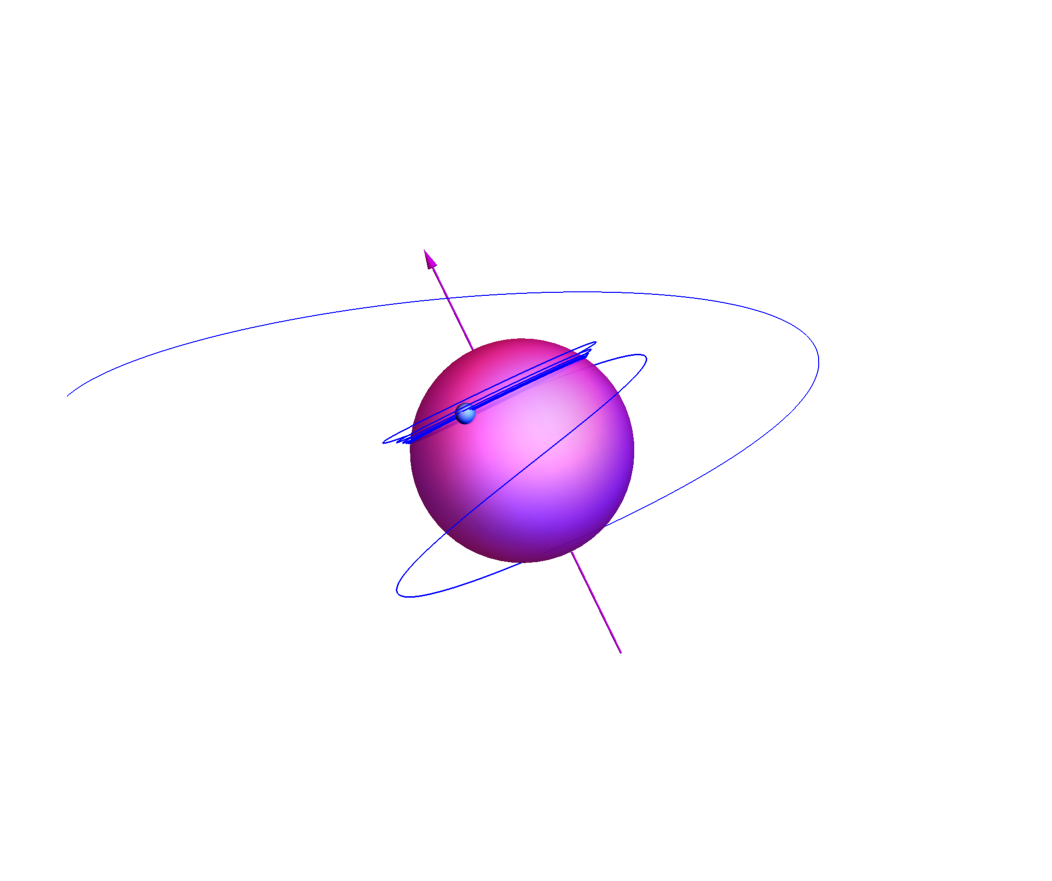} 
	\hfill
	\includegraphics[angle=0,width=0.4\textwidth]{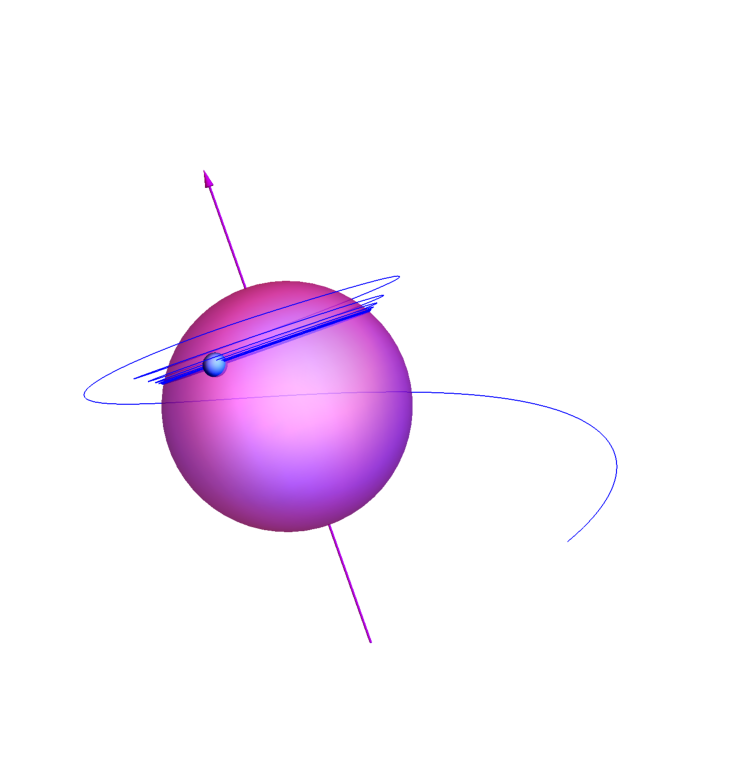} 
	\caption{{\bf Left panel}: Trajectory of the test planet with parameters $\gamma=0.85$, $\lambda=1.7$ and $Q=1$, which is plunging into the fast-rotating Kerr black hole with spin $a=0.9982$. This planet is winding up on the black hole event horizon higher the equatorial plane. {\color{blue}Blue} curve here and in the Fig.~\ref{fig6} is the numerically calculated geodesic trajectory with using equations of motion (\ref{dr})--(\ref{dt}) for massless test particles like photons ($\mu=0$). {\bf Right panel}: Trajectory of the test planet with parameters $\gamma=0.85$, $\lambda=1.7$ and $Q=1$, which is plunging into the fast-rotating Kerr black hole with spin $a=0.9982$. This planet is winding up on the black hole event horizon higher the equatorial plane. {\color{blue}Blue} curve here and in the Fig.~\ref{fig6} is the numerically calculated geodesic trajectory with using equations of motion (\ref{dr})--(\ref{dt}) for massless test particles like photons ($\mu=0$).}
	\label{fig5}
\end{figure}

\begin{figure}
	\centering
	\includegraphics[angle=0,width=0.41\textwidth]{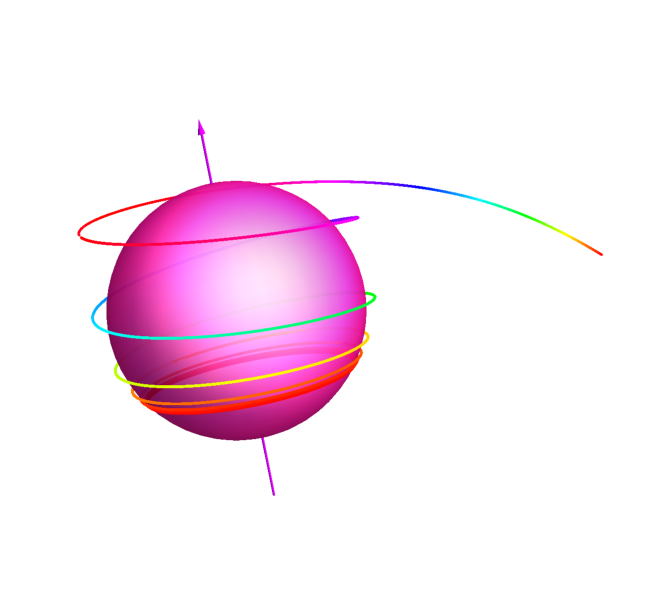} 
	\hfill
	\includegraphics[angle=0,width=0.56\textwidth]{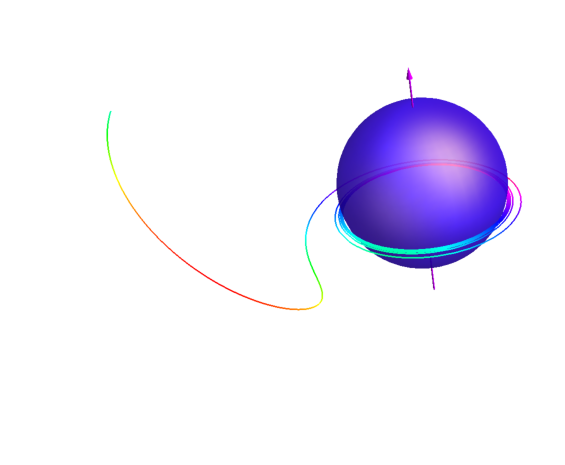} 
	\caption{{\bf Left panel}: Trajectory of the photon with parameters $\lambda=2$ and $Q=1$, which is plunging into the fast-rotating Kerr black hole with $a=0.9982$ and is winding up on the black hole event horizon below the equatorial plane. {\color{blue}Multi}--{\color{red}colored} curve is the numerically calculated geodesic trajectory with using equations of motion (\ref{dr})--(\ref{dt}) for massless test particles like photons ($\mu=0$). {\bf Right panel}: Trajectory of the photon with parameters $\lambda=-6.5$ and $Q=4$, which is plunging into the fast-rotating Kerr black hole with $a=0.9982$ and is winding up on the black hole event horizon below the equatorial plane.}
	\label{fig6}
\end{figure}

\begin{figure}
	\centering
	\includegraphics[angle=0,width=0.45\textwidth]{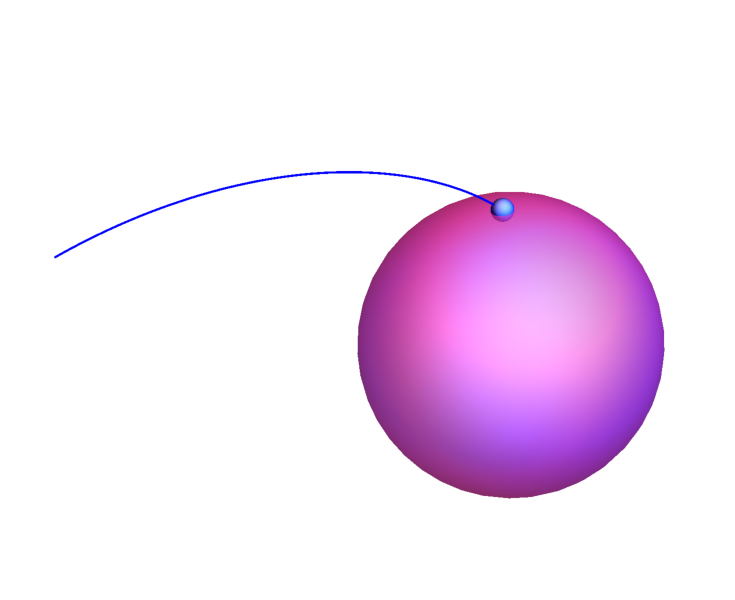} 
	\caption{A trivial but though very expressive numerically calculated trajectory of a test planet ($\mu\neq0$) with parameters $\gamma=1$, $\lambda=1$ and $Q=0.5$, which is plunging into the spherically {\it symmetric} and nonrotating Schwarzschild black hole (with both the spin $a=0$ and electric charge $q=0$). The starting point for this crazy voyage is at the radial distance $r=6$.}
	\label{fig7}
\end{figure}

We start to elucidate the versatility of intriguing symmetry and asymmetry phenomena outside and inside black holes by using the Carter--Penrose diagrams (for details see, i.\, e., \cite{mtw,Chandra,Wald,hawkingellis}), describing in particular the global space-time structure of black holes. The evident manifestation of {\it symmetry} of this global structure is infinite space volumes as outside and inside the black hole event horizon. See in Fig.~\ref{fig1}  the corresponding Carter--Penrose diagram for the Reissner--Nordstr\"om black hole, which is a special spherically {\it symmetric} case of Kerr-Newman black hole without rotation i.\, e., $a=0$ but $q\neq0$. From the pure geometric point of view this diagram is both left-right and up-down symmetric. On the contrary, from the physical or space-time point view this diagram is absolutely {\it asymmetric} due to the inexorable upward flow of time not only at this diagram but throughout the whole universe. More precisely it means that in the General Relativity all objects are allowed to move only inside the upward directed light cones (at $\pm45^\circ$ with respect to the upward direction. The upward directed light cone is the inexorable {\it asymmetry} of the world.

$2D$ presentation of the voyage through interiors of Reissner--Nordstr\"om black hole by using the Einstein--Rosen bridge is shown in Fig.~\ref{fig2}. The electric charge of the black hole is $q=0.99$. A test planet (or spaceship) with the electric charge $\epsilon=-1.5$ is periodically orbiting around black holes with orbital parameters $\gamma=0.5$, $\lambda=0.5$, corresponding to the maximal radius (apogee) $r_{\rm max}=1.65$ and minimal radius (perigee) $r_{\rm min}=0.29$, respectively, in dimensional units. 

The periodic planet geodesic trajectories ({\color{magenta}magenta} curves, both 2 Fig.~\ref{fig2} and Fig.~\ref{fig4}), were calculated  numerically with using equations of motion (\ref{dr})--(\ref{dt}) for massive test particles ($\mu\neq0$). Note, that the periodical motion of the test planet is limited in time due to energy losses in inevitable emission of the gravitational waves. 

The picture in Fig.~\ref{fig2} is geometrically absolutely {\it symmetric} or, in other words, it is completely or nicely {\it symmetric}. The geodesic trajectories of test planet ($\mu\neq0$) in this Fig.~\ref{fig2} and in Fig.~\ref{fig3} (the {\it \color{red}red}-colored curves), are numerically calculated \cite{Babichev13,Dokuch14, FizLab, DokEr15,doknaz17, doknaz18b, doknaz19, doknazsm19,doknaz19b, dokuch19,doknaz18c}, by using the corresponding equations of  motion in the Kerr-Newman metric (\ref{dr})--(\ref{dt}).

At the same time, this picture is misleading and physically controversial: Indeed, the voyage is starting at apogee $r_a$ from the position at point ${\color{magenta}A}$, then reach the perigee $r_p$ and return the  apogee $r_a$ at the point ${\color{magenta}B}$ for a finite proper time, demonstrating the absolute geometric {\it symmetry}. Meanwhile, there is a crucial hitch: this apogee $r_a$ at the point ${\color{magenta}B}$ is not in the native universe, but in the other quite distant universe, as it is clearly viewed at the Carter--Penrose diagram in Fig.~\ref{fig1}. This hitch again destroys the Einstein--Rosen bridge {\it symmetry}.

Figure~\ref{fig3} shows the embedding diagram for the voyage through black hole interiors by using the Einstein--Rosen bridge. The embedding diagram is very useful for the training of intuitive understanding of the peculiarities of the enigmatic black holes. In this embedding diagram the Einstein--Rosen bridge connects two asymptotically flat universes like wormhole tunnel \cite{Wheeler62,MorrisThorne}, but with the only one-way motion from the initial point ${\color{magenta}A}$ to the final point ${\color{magenta}B}$. The geometrical {\it symmetry} of this embedding diagram is deceptive. In fact, this embedding diagram demonstrate the {\it asymmetric} space-time origin of the one-way Einstein--Rosen bridge (remember about loss-cone).

The completely {\it symmetric} picture of the periodic orbital motion of the test planet or spaceship around the central singularity of the Reissner--Nordstr\"om black hole inside the Cauchy horizon $r_-$ is shown in Fig.~\ref{fig4}. The electric charge of the black hole is $q=0.99$ and test planet (or spaceship) with the electric charge $\epsilon=-1.5$ is periodically orbiting around black holes with orbital parameters $\gamma=1.7$, $\lambda=0.1$, corresponding to the maximal radius (apogee) $r_{\rm max}=0.75$ and minimal radius (perigee) $r_{\rm man}=0.09$, respectively. The {\it asymmetric} Reissner--Nordstr\"om bridge is only needed for penetration into this very exotic region at $0<r<r_-$, where exist the nearly stable periodic orbits for test particles \cite{BichakSB89,BBS89,GruKagr11, OSLV,HKKL, PugQueRuff, Dokuch11}, which are very similar to the periodic orbits outside the black hole event horizon $r_+$.

\section{Symmetry and asymmetry of test particle trajectories near rotating black hole}

Figures~\ref{fig5}--\ref{fig7} demonstrate both {\it symmetry} and {\it asymmetry} features of massive and massless particle trajectories plunging into rotating Kerr black hole with spin $a=0.9982$. {\color{magenta}Magenta} arrows show the direction of the black hole rotation in accordance with the gimlet rule. {\color{blue}Multi}--{\color{red}colored} curves at Figs.~\ref{fig7} and \ref{fig7} are the geodesic trajectories for massless test particles like photons ($\mu=0$) numerically calculated with using equations of motion (\ref{dr})--(\ref{dt}). By approaching the black hole, the trajectories of all particles, both massive and massless ones, are infinitely winding up on the black hole event horizon in the direction of black hole rotation and at the fixed latitudes. This winding up is a {\it symmetry} manifestation of all trajectories behavior, which are plunging into rotating black hole. At the same direction of black hole rotation is a corresponding  {\it asymmetry} manifestation of the gravitational field of the Kerr metric.

At last, for completeness of black hole {\it symmetric} and {\it asymmetric} properties, at Figs.~\ref{fig8} is shown the trajectory of the test planet ($\mu\neq0$) with parameters $\gamma=1$, $\lambda=1$ and $Q=0.5$. This test planet is plunging into the spherically {\it symmetric} and nonrotating Schwarzschild black hole (with both the spin $a=0$ and electric charge $q=0$), starting from the radial distance $r=6$.

It must be especially checked that the traversable (though only one-way in time and direction) Einstein--Rosen bridge is absent at all inside the Schwarzschild black hole (see for details, e.\, g., \cite{mtw, Penrose}).

\section{Conclusion and Discussion}

It is demonstrated the symmetry and asymmetry of the voyage on one-way Einstein--Rosen bridge inside black hole toward the endless multiplanetary future. 

The apparent symmetry of both the Carter--Penrose and embedding diagrams is mainly related with a pure geometrical vision of this phenomenon. Quite the contrary, the physical (space-time) vision elucidates the absolute asymmetry of the Einstein--Rosen bridge due to existence of the light-cone limitation for possible motions.  

Note, that the traversable (though only one-way in time and direction) Einstein--Rosen bridge exist only in the case of both rotating Kerr $a\neq0$ and electrically charged Reissner--Nordstr\"om $q\neq0$ black holes. It is absent at all inside the Schwarzschild black hole (see for details, e.\, g., \cite{mtw, Penrose}).

The infinite winding up of particle trajectories on the black hole event horizon is a manifestation of {\it symmetry} behavior of all trajectories, plunging into rotating black hole. At the same time, the fixed direction in space of the black hole rotation axis is a strict manifestation of the Kerr metric both {\it asymmetry} and {\it asymmetry}. 

We are grateful to E. O. Babichev, V. A. Berezin, Yu. N. Eroshenko, N. O. Nazarova and A. L. Smirnov for stimulating discussions. Authors also are very indebted to anonymous reviewer for suitable references and historical remarks which improve the presentation of paper.


\begin{thebibliography}{99}

\bibitem{BoyerLindquist} R. H. Boyer and R. W. Lindquist, J. Math. Phys. 8, 265 (1967).

\bibitem{Carroll} S. Carroll, An Introduction to General Relativity, new international edition, Pearson (2014), p. 257.

\bibitem{Ullmann} V. Ullmann, Gravity, Black Holes and the Physics
of Time-Space; Czechoslovak Astronomic Society, CSAV, Ostrava, (Online version in English: https://astronuclphysics.info/GravitCerneDiry.htm) (1986).

\bibitem{Giirsel} Y. Giirsel, V. D. Sandberg, I. D. Novikov, and
A.A. Starobinskij, Phys. Rev. D 19, 413 (1979).

\bibitem{SimpsonPenrose} M. Simpson and R. Penrose, Int. J. Theor. Phys. 7, 183 (1973).

\bibitem{DeMott} R. DeMott, S. DeMott, and A. DeMott, Class. Quant.
Grav. 39, 195015 (2022).

\bibitem{Abramson} D. Abramson, Thai J. of Phys. 38, 69 (2021).

\bibitem{DysonMeent} C. Dyson and M. van de Meent, Class. Quant. Grav.
40, 195026 (2023).

\bibitem{Kerr} Kerr, R.P. Gravitational field of a spinning mass as an example of algebraically special metrics  {\em Phys. Rev. Lett.} {\bf 1963}, {\em 11}, 237--238.

\bibitem{Carter68} Carter, B. Global Structure of the Kerr Family of Gravitational Fields. {\em Phys. Rev.} {\bf 1968}, {\bf 174}, 1559--1570.

\bibitem{novthorne73} Novikov I.D, Thorne K.S. in {\em Black Holes} (Eds C DeWitt, B S DeWitt) (New York: Gordon and Breach, 1973) p. 343--350.

\bibitem{mtw} Misner, C.W.; Thorne, K.S.; Wheeler J.A. {\em Gravitation}; W. H. Freeman: San Francisco, CA, USA, 1973.

\bibitem{Chandra} Chandrasekhar, S. The Mathematical Theory of Black Holes. In \emph{The International Series of Monograph on Physics}; Clarendon Press: Oxford, UK, 1983; Chapter 7, Volume 69.


\bibitem{Penrose} R. Penrose Structure of space-time/ Battelle Rencontres 1967. Lectures in Mathematical Physics Chapter VII. Editted by Cecille M. Dewitt and John A, Wheeler. (W. A.Benjamin Inc. New York-Amsterdam 1968) Chapter 2.

\bibitem{Bardeen73} Bardeen, J.M. {\em Black Holes}; DeWitt, C., DeWitt, B. S., Eds.; Gordon and Breach Science Publishers:  New~York, NY,~USA, 1973; pp. 215--239.

\bibitem{Choquet-Bruhat} Yvonne Choquet-Bruhat, Cecile DeWitt-Morette, Margarete Dillard-Bleick, Analysis, Manifolds and Physics, Part 1: Basics (Elsevier Science B. V. Amsterdam, The netherland, 1977), Chapter~V.

\bibitem{Wald} Wald R M {\em General Relativity} (Chicago, IL: The Univ. of Chicago Press, 1984).

\bibitem{hawkingellis} Hawking S.W.; Ellis, G.F.R. {\em The Large--Scale Structure of Space--Time} (Cambridge Monographs on Mathematical Physics Cambridge: University Press, 2011).

\bibitem{BPT} Bardeen, J.M.; Press, W.H.; Teukolsky, S.A.  Rotating Black Holes: Locally Nonrotating Frames, Energy Extraction, and Scalar Synchrotron Radiation.  {\em Astrophys.~J.} {\bf 1972}, \emph{178}, 347--370. 

\bibitem{bch} Bardeen, J.M.; Carter, B.; Hawking S.W. The four laws of black hole mechanics. {\em Commun. Math. Phys.} {\bf 31} 161--170. (1973)

\bibitem{Kling} T. P. Kling, E. Grotzke, K. Roebuck, and H.Roebuck,
Gen. Rel. Grav. 51, 32 (2019).

\bibitem{Babichev13} Babichev, E.O.; Dokuchaev, V.I.; Eroshenko, Y.N.  Black holes in the presence of dark energy. {\em Phys. Usp.} {\bf 2013}, \emph{56}, 1155--1175.

\bibitem{Dokuch14} Dokuchaev, V.I. Spin and mass of the nearest supermassive black hole. {\em Gen. Relat. Gravit.} {\bf 2014}, \emph{46}, 1832--1845.

\bibitem{FizLab} Dokuchaev, V.I.; Eroshenko, Yu. N. Physical laboratory at the center of the Galaxy. {\em Phys. Usp.} {\bf 2015}, \emph{58}, 772--784.

\bibitem{DokEr15} Dokuchaev, V.I.; Eroshenko, Yu. N. Weighing of the dark matter at the center of the Galaxy.  {\em JETP Lett.} {\bf 2015}, \emph{101}, 777--782.

\bibitem{doknaz17} Dokuchaev, V.I.; Nazarova, N.O. Gravitational lensing of a star by a rotating black hole. {\em J. High Energy Phys.~Lett.} {\bf 2017}, \emph{106},  637--642.

\bibitem{doknaz18b} Dokuchaev, V.I.; Nazarova, N.O. Star Motion Around Rotating Black Hole. {\bf 2017},  https://youtu.be/P6DneV0vk7U.

\bibitem{doknaz19} Dokuchaev, V.I.; Nazarova, N.O. Event horizon image within black hole shadow. {\em J. Exp. Theor. Phys.} {\bf 2019}, \emph{128}, 578--585. 

\bibitem{doknazsm19} Dokuchaev, V.I.; Nazarova, N.O.; Smirnov, V.P. Event horizon silhouette: Implications to supermassive black holes in the galaxies M87 and Milky Way. {\em Gen. Relat. Gravit.} {\bf 2019}, \emph{51},~81. 

\bibitem{doknaz19b} Dokuchaev, V.I.; Nazarova, N.O. The~Brightest Point in Accretion Disk and Black Hole Spin: Implication to the Image of Black Hole M87*. {\em Universe} {\bf 2019}, \emph{5}, 183. 

\bibitem{dokuch19} Dokuchaev, V.I. To see the invisible: Image of the event horizon within the black hole shadow.  {\em Int. J. Mod. Phys. D} {\bf 2019}, \emph{28}, 1941005.

\bibitem{doknaz18c} Dokuchaev, V.I.; Nazarova, N.O. Infall of the star into rotating black hole viewed by a distant observer. {\bf 2019}, https://youtu.be/fps-3frL0AM.

\bibitem{Wheeler62} Wheeler, J.A. \textit{Geometrodynamics}, Academic Press: New York, NY, USA, 1962; 334p.

\bibitem{MorrisThorne} Morris, M.S.; Thorne, K.S. Wormholes in spacetime and their use for interstellar travel: a tool for teaching general relativity.  {\em Am. J. Phys.} {\bf 1988}, \emph{56}, 395-412.

\bibitem{BichakSB89} Bi\v c\'ak, J.; Stuchl\'ik,  Z.; Balek V. The motion of the charged particles in the field of rotating charged black holes and naked singularities I. The general features of the radial motion and the motion along the axis of symmetry.  {\em Bull. Astron. Inst. Czech.} {\bf 1989}, \emph{40}, 65--92.

\bibitem{BBS89} Balek, V.; Bi\v c\'ak, J.; Stuchl\'ik, Z. The motion of the charged particles in the field of rotating charged black holes and naked singularities II. The motion
in the equatorial plane.  {\em Bull. Astron. Inst. Czech.} {\bf 1989}, \emph{40},  133-165.

\bibitem{GruKagr11} Grunau, S.; Kagramanova, V. Geodesics of electrically and magnetically charged test particles in the Reissner--Nordstr\"om spacetime: analytical solutions. {\em Phys. Rev. D} {\bf 2011}, \emph{83}, 044009, 18pp.

\bibitem{OSLV} Olivares. M.; Saavedra, J.; Leiva, C.; Villanueva, J.R. Motion of charged particles on the Reissner--Nordstr\"om (anti)-de Sitter black holes.  {\em Mod. Phys. Lett. A} {\bf 2011}, \emph{26}, 2923--2950.

\bibitem{HKKL} Hackmann, E.; Kagramanova, V.; Kunz, J.; Lam\"merzahl, C. Analytical solution of the geodesic equation in Kerr-(anti-) de Sitter space-times.  {\em Phys. Rev. D} {\bf 2010}, \emph{81}, 044020, 18pp.

\bibitem{PugQueRuff} Pugliese, D.; Quevedo, H.; Ruffini, R. Circular motion of neutral test particles in Reissner--Nordstr\"om spacetime.  {\em Phys. Rev. D} {\bf 2011}, \emph{83}, 024021, 23pp.

\bibitem{Dokuch11} Dokuchaev, V.I. Is there life inside black holes? {\em Class. Quantum Grav.} {\bf 2008}, {\em 28}, 235015, pp10.

\end{thebibliography}
\end{document}